\documentclass[lettersize,journal]{IEEEtran}
\usepackage{threeparttable}

\usepackage{algorithmic}
\usepackage{algorithm}
\usepackage{array}
\usepackage{textcomp}
\usepackage{stfloats}
\usepackage{url}
\usepackage{verbatim}
\usepackage{graphicx}
\usepackage{cite}
\usepackage{amsmath,amsfonts,amssymb}
\usepackage{booktabs}
\usepackage{tikz}
\usepackage{hyperref}
\usepackage{geometry}
\usepackage{graphicx}
\usepackage[caption=false,font=normalsize,font=sf,textfont=sf]{subfig}

\usepackage{makecell}
\definecolor{starcolor}{RGB}{0,150,0}
\definecolor{rhythmcolor}{RGB}{0,90,200}
\usetikzlibrary{calc}
\usetikzlibrary{arrows.meta, positioning}

\begin{document}

\title{Confidence-Guided Markov Weighting for Semi-Supervised Tabla Stroke Transcription}

\author{Rahul Bapusaheb Kodag, Vipul Arora 
\thanks{This paper was produced by the IEEE Publication Technology Group. They are in Piscataway, NJ.}
\thanks{Manuscript received April 19, 2021; revised August 16, 2021.}}

\markboth{Journal of \LaTeX\ Class Files,~Vol.~14, No.~8, August~2021}%
{Shell \MakeLowercase{\textit{et al.}}: A Sample Article Using IEEEtran.cls for IEEE Journals}


\maketitle

\begin{abstract}
Tabla Stroke Transcription (TST) converts tabla audio into symbolic stroke sequences, but the scarcity of annotated recordings makes fully supervised training challenging. We propose a semi-supervised framework that uses sequence-level labelled and unlabelled tabla recordings. A teacher model generates pseudo-label sequences, while a Stroke-Level Confidence Estimation Model (S-CEM) estimates confidence for each predicted stroke. To improve training with uncertain pseudo-labels, we propose Confidence-Guided Markov Weighted Alternative Temporal Classification (CMW-ATC), which weights candidate sequences within uncertain spans using learned stroke transitions and reliable neighbouring strokes. Experiments across three TST evaluation settings show consistent improvements over conventional teacher--student training and Alternative Temporal Classification in its replacement form (ATC-R). Ablation studies further show the contributions of S-CEM, Markov transition weighting, and the use of reliable strokes on both sides of uncertain spans.
\end{abstract}

    
\section{Introduction}

Tabla Stroke Transcription (TST) converts tabla audio into symbolic stroke
sequences, supporting applications such as $t\bar{a}la$ identification,
rhythmic-pattern analysis, pedagogy, and digital preservation. Developing
accurate TST systems remains challenging because of the scarcity of annotated
data. Many supervised approaches require precise onset-level annotations,
which are expensive and time-consuming to obtain and typically require expert
tabla players.

Previous TST studies have explored onset-based stroke segmentation and
classification~\cite{S12,1_gowriprasad2020onset}, as well as strongly
supervised deep-learning approaches~\cite{S9,S10}. Meta-learning has further
enabled TST with limited onset-level annotated data~\cite{meta}. The Adaptive
Dynamic Rhythm Language Model (ADRM) enabled TST from sequence-level stroke
annotations without explicit onset timings~\cite{ADRM}. Although this reduces
the annotation detail required for TST, sequence-level stroke annotations are
still required for training.

Unsupervised learning can further reduce dependence on manual annotation.
Recent work has used acoustic clustering to learn stroke representations
across tabla and other Indian percussion instruments without predefined
stroke labels~\cite{1_4_gowriprasad2025unsupervised}. Beyond purely
unsupervised approaches, unlabelled recordings can also be incorporated
through semi-supervised learning, where the available sequence-level
annotations continue to provide supervision. In this setting, a teacher
model generates pseudo-labels for the unlabelled recordings, but errors in
these sequences can propagate to the student model.

Connectionist Temporal Classification (CTC)~\cite{CTC} enables training from
sequence-level labels without explicit frame-level alignment, but assumes
that the supplied target sequence is correct. Graph-based extensions have
relaxed this assumption by allowing alternative supervision structures
~\cite{STC,W-CTC,BTC,OTC,GTC}. Alternative Temporal Classification (ATC)~\cite{ATC}
further uses pseudo-label confidence to relax uncertain positions during
semi-supervised training. In its replacement form, ATC-R assigns the same
alternative weight to all non-blank labels at an uncertain position and does
not exploit the sequential relationships among stroke labels.

This limitation is particularly relevant in situations where multiple
consecutive pseudo-labels are uncertain, since an uncertain stroke can
provide unreliable context for the next position. A conventional Markov
model uses forward transitions conditioned on the preceding stroke,
whereas a Markov bridge conditions the intermediate path on both endpoint
states~\cite{MarkovBridge,MarkovBridgeInference}. To the best of our
knowledge, this principle has not been used for uncertainty-aware
pseudo-label supervision in speech or music transcription. Motivated by
this principle, we propose Confidence-Guided Markov Weighted Alternative
Temporal Classification (CMW-ATC), which uses learned stroke transitions
and the available reliable boundary information to weight candidate
sequences within uncertain spans.

Identifying which pseudo-label strokes should be treated as uncertain also
requires reliable stroke-level confidence estimates. Existing confidence estimation methods
have primarily been developed for words or subwords in ASR
~\cite{ravi2025asr}, whereas tabla strokes are atomic prediction classes.
We therefore develop a Stroke-Level Confidence Estimation Model (S-CEM) that
predicts the confidence that each teacher-generated stroke is correct from
teacher-derived representations under sequence-level supervision.
The main contributions of this work are:
\begin{enumerate}
\item We introduce CMW-ATC for uncertainty-aware semi-supervised
pseudo-label training. It forms candidate stroke sequences within uncertain
spans and assigns them weights using learned stroke transitions and available
reliable boundaries.

\item We introduce S-CEM for stroke-level confidence estimation. It
estimates the correctness of each teacher-predicted stroke using acoustic
representations and CTC posterior information.

\item We conduct component ablation studies across three TST evaluation
settings. The experiments separately examine the contributions of S-CEM,
forward Markov weighting, and right-boundary conditioning.

\end{enumerate}

The remainder of this paper is organized as follows.
Section~\ref{sec:literature} reviews related work.
Section~\ref{sec:proposed_method} presents the proposed methodology.
Section~\ref{sec:exp_setup} describes the experimental setup, followed by
results and discussion in Section~\ref{sec:results}. Finally,
Section~\ref{sec:conclusion} concludes the paper.

\section{Related Work}
\label{sec:literature}

\subsection{Tabla Stroke Transcription}

Research on Tabla Stroke Transcription (TST) has explored acoustic stroke
characterization and automatic recognition from tabla recordings. Early
approaches followed a segment-and-classify paradigm using acoustic features
and statistical models~\cite{S2,S12}. Subsequent studies investigated the
acoustic characteristics of tabla strokes~\cite{S12_1}, temporal relationships
between successive strokes~\cite{S8}, and onset detection for identifying
stroke boundaries~\cite{1_gowriprasad2020onset}. Related work on mridangam has
also explored acoustic and statistical approaches to stroke transcription
~\cite{G1,G2,G3}.

More recent TST approaches have adopted supervised deep-learning models,
including models adapted from automatic drum transcription~\cite{S9} and
transfer-learning approaches using Western drum data~\cite{S10}. These
approaches rely on onset-level stroke annotations. Meta-learning has further
improved data efficiency by enabling TST with a limited amount of
onset-annotated data~\cite{meta}, but still requires temporal annotations.
Our previous work introduced the Adaptive Dynamic Rhythm Language Model
(ADRM), which enabled sequence-level TST without explicit onset timings and
incorporated rhythmic information during CTC-lattice decoding~\cite{ADRM}.
However, ADRM still requires sequence-level stroke annotations for training.

Unsupervised learning provides another direction for reducing dependence on
manual annotation. Recent work has used timbre-based clustering to learn
stroke representations across tabla and mridangam without predefined stroke
labels~\cite{1_4_gowriprasad2025unsupervised}. The resulting clusters can subsequently be mapped to the target stroke
vocabulary using labelled data.

\subsection{Graph-Based Extensions of CTC}

Connectionist Temporal Classification (CTC)~\cite{CTC} provides an
alignment-free objective for sequence-level supervision and is therefore
well suited to TST when only ordered stroke sequences are available.
However, standard CTC assumes that the supplied target sequence is correct.
Several extensions modify the CTC training graph to handle incomplete or
erroneous supervision. W-CTC~\cite{W-CTC} introduces wildcard transitions
for partial label sequences, while STC~\cite{STC} allows missing labels at
arbitrary positions. BTC~\cite{BTC} introduces bypass transitions for
imperfect transcripts, and OTC~\cite{OTC} combines wildcard and bypass
mechanisms to accommodate different transcription errors.

Graph-based Temporal Classification (GTC)~\cite{GTC} generalizes CTC to
weighted graph supervision, allowing multiple candidate target sequences and
their weights to be marginalized during training. In semi-supervised ASR,
GTC constructs supervision graphs from multiple pseudo-label hypotheses and
their associated scores, providing a general framework for weighted
alternative supervision.

Markov models provide a standard framework for modeling state transitions,
whereas a Markov bridge describes a Markov path conditioned on specified
endpoint states~\cite{MarkovBridge}. Such endpoint-conditioned paths have
also been used for statistical inference in Markov processes
~\cite{MarkovBridgeInference}. This formulation is relevant to uncertain
sequence spans because reliable states before and after an uncertain region
can provide additional information for weighting the possible intermediate
sequences.

Alternative Temporal Classification (ATC)~\cite{ATC} uses pseudo-label
confidence to identify uncertain tokens during semi-supervised training.
Its ATC-R variant replaces a low-confidence token with a wildcard that
accepts any non-blank label, but does not use the sequential relationships
among these alternatives when assigning their weights.

These lines of work provide the basis for our approach. ATC provides
confidence-guided alternative supervision, while Markov-bridge conditioning
provides a way to use reliable neighbouring states when weighting uncertain
intermediate sequences. CMW-ATC combines these principles by using learned
stroke transitions and the available reliable boundary information to weight
candidate sequences within uncertain pseudo-label spans.

\subsection{Confidence Estimation}
\label{sec:confidence_related}

Confidence estimation is widely used to assess the reliability of model
predictions. Posterior probability and entropy provide simple confidence
measures, but neural models can produce overconfident predictions
~\cite{Confidence_survey,corbiere2019addressing}. Auxiliary Confidence
Estimation Models (CEMs) instead learn confidence from model-derived
representations. In ASR, CEMs have been trained using binary correctness
targets obtained by aligning hypotheses with reference sequences
~\cite{CEM_ASR}. TeLeS~\cite{ravi2024teles} introduced continuous targets
that combine temporal and lexical similarity, but requires temporal
alignment between hypothesis and reference. TruCLeS~\cite{ravi2025asr}
avoids this temporal-alignment requirement by constructing continuous
targets from true-class probabilities and lexical similarity.

These approaches were developed for ASR, where outputs are words or
subwords and lexical similarity can represent partial correctness. In TST,
each tabla stroke is an atomic prediction class, so lexical similarity
between stroke labels does not provide a meaningful measure of
transcription correctness. We therefore use a task-specific stroke-level
CEM with binary correctness targets under sequence-level supervision.

\section{Proposed Method}
\label{sec:proposed_method}
The proposed framework consists of teacher-generated pseudo-labels,
stroke-level confidence estimation, and confidence-guided student training.
The student is trained using conventional Connectionist Temporal
Classification (CTC) on labelled recordings and the proposed
Confidence-Guided Markov-Weighted Alternative Temporal Classification
(CMW-ATC) objective on pseudo-labelled recordings. CMW-ATC uses stroke-level confidence and learned stroke transitions to
guide training on pseudo-labelled recordings.

Let \(X\in\mathbb{R}^{F\times T}\) denote an input acoustic feature matrix consisting of \(F\) feature coefficients over \(T\) frames, and let \(Y=[y_1,\ldots,y_U]\in\mathcal{S}^{U}\) denote a corresponding stroke sequence of length \(U\), where \(y_u\in\mathcal{S}\) and \(\mathcal{S}=\{s_1,s_2,\ldots,s_K\}\) is the tabla stroke vocabulary. The labelled and unlabelled datasets are denoted by \(D_L=\{(X_i,Y_i)\}_{i=1}^{N}\) and \(D_U=\{X_j\}_{j=1}^{M}\), respectively, where \(N\) and \(M\) denote the numbers of labelled and unlabelled recordings.

\subsection{CTC Preliminaries}
Connectionist Temporal Classification (CTC) enables sequence prediction without
requiring explicit frame-level alignment between the input audio and target
stroke sequence. It introduces a blank symbol and marginalizes over all
frame-level paths that collapse to the target sequence.
Let $\bar{\mathcal{S}}=\mathcal{S}\cup\{\varnothing\}$ denote the CTC output
label set, where $\varnothing$ is the blank symbol, and let $\mathcal{B}$ denote
the CTC collapse function that merges consecutive
repetitions and removes blanks. For input $X$ and target stroke sequence
$Y=[y_1,\ldots,y_U]$, the CTC probability is
\begin{equation}
P_{\theta}^{\mathrm{CTC}}(Y\mid X)=
\sum_{\pi:\mathcal{B}(\pi)=Y}
\prod_{t=1}^{T}
p_{\theta}(\pi_t\mid X,t),
\label{eq:ctc_probability}
\end{equation}
where $\pi\in\bar{\mathcal{S}}^{T}$ is a frame-level CTC path. The corresponding
training loss is
\begin{equation}
\mathcal{L}_{\mathrm{CTC}}(X,Y)=
-\log P_{\theta}^{\mathrm{CTC}}(Y\mid X).
\label{eq:ctc_loss}
\end{equation}
This CTC formulation is used for labelled recordings, while CMW-ATC extends
it to pseudo-labelled recordings by marginalizing over alternative stroke
sequences at uncertain positions.

\subsection{Teacher Model and Pseudo-Label Generation}

A teacher acoustic model with parameters $\theta_T$ is trained on the labelled
dataset using the CTC loss in Eq.~\eqref{eq:ctc_loss}.
After training, the teacher is frozen and used to generate pseudo-label
sequences for the unlabelled recordings. For an input $X$, the teacher produces a pseudo-label sequence
$\hat{Y}=[\hat{y}_1,\ldots,\hat{y}_{\hat{U}}]$ using CTC decoding, where
$\hat{U}$ is the length of the decoded sequence. The stroke-transition model is not used during teacher decoding, so
pseudo-labels are generated only by the acoustic model.

\subsection{Stroke-Level Confidence Estimation}
\label{sec:confidence}
Pseudo-labelled sequences may contain stroke predictions with different
levels of reliability. We therefore adapt the Confidence Estimation Model
(CEM) framework of TruCLeS~\cite{ravi2025asr} to develop a Stroke-Level
Confidence Estimation Model (S-CEM) for TST. The S-CEM uses teacher representations for each predicted stroke to estimate
stroke-level confidence.

Unlike ASR words, tabla strokes are discrete stroke labels. Consequently,
lexical similarity between stroke labels does not provide a meaningful
measure of partial correctness. We therefore replace the continuous target used by TruCLeS, which includes
lexical similarity, with a binary target indicating whether each predicted
stroke is correct.

For S-CEM training, the same frozen teacher is applied to the labelled
training recordings to obtain predicted stroke sequences and teacher
representations. For an input recording $X$, let
$\hat{Y}=[\hat{y}_1,\ldots,\hat{y}_{\hat{U}}]$ denote the teacher
prediction and $Y=[y_1,\ldots,y_U]$ the reference stroke sequence. The two
sequences are aligned using stroke-level Levenshtein alignment. For each
predicted stroke $\hat{y}_u$, the binary correctness target is defined as

\begin{equation}
b_u =
\begin{cases}
1, & \text{if $\hat{y}_u$ is matched to the same reference stroke},\\
0, & \text{otherwise}.
\end{cases}
\label{eq:stroke_correctness}
\end{equation}
A deletion does not produce
a predicted stroke and therefore has no corresponding stroke-level
target. The Levenshtein alignment is used only to determine the
correctness targets and does not require onset-level annotations.

To obtain a teacher representation for each predicted stroke occurrence,
we compute the most probable CTC alignment. The
constrained Viterbi alignment is
\begin{equation}
\pi^{*}
=
\arg\max_{\substack{
\pi\in\bar{\mathcal{S}}^{T}\\
\mathcal{B}(\pi)=\hat{Y}}}
P_{\theta_T}(\pi\mid X),
\label{eq:teacher_viterbi}
\end{equation}
where $\theta_T$ denotes the frozen teacher parameters. Each predicted stroke occurrence is treated separately in the alignment, so
repeated occurrences of the same stroke label remain distinct.

Let $\mathcal{T}_u$ denote the non-blank output indices assigned by
$\pi^{*}$ to the $u$th predicted stroke occurrence. Let
$r_t\in\mathbb{R}^{D}$ denote the teacher representation immediately
before the CTC output layer, and let
$p_t\in\mathbb{R}^{K+1}$ denote the corresponding CTC posterior vector,
where the additional dimension corresponds to the CTC blank. The
stroke-level representations are obtained by temporal averaging:
\begin{equation}
\bar{r}_u =
\frac{1}{|\mathcal{T}_u|}
\sum_{t\in\mathcal{T}_u} r_t,
\qquad
\bar{p}_u =
\frac{1}{|\mathcal{T}_u|}
\sum_{t\in\mathcal{T}_u} p_t.
\label{eq:stroke_features}
\end{equation}
The resulting stroke-level feature vector is
\begin{equation}
\phi_u=[\bar{r}_u;\bar{p}_u]
\in\mathbb{R}^{D+K+1},
\label{eq:stroke_evidence}
\end{equation}
which is used as input to the S-CEM. The S-CEM produces a continuous
stroke-level confidence score
\begin{equation}
c_u =
S\text{-}CEM_{\theta_c}(\phi_u),
\qquad
c_u\in(0,1).
\label{eq:cem_output}
\end{equation}
The S-CEM is trained using binary cross-entropy:
\begin{equation}
\mathcal{L}_{\mathrm{S\text{-}CEM}}
=
-\sum_{u=1}^{\hat{U}}
\left[
b_u\log c_u
+
(1-b_u)\log(1-c_u)
\right]
\label{eq:cem_loss}
\end{equation}
Higher $c_u$ indicates greater confidence that the predicted stroke is
correct.

For an unlabelled recording, the teacher prediction and constrained CTC
alignment provide the stroke-level feature vectors $\phi_u$, from which
the trained S-CEM produces a confidence score $c_u$ for each predicted
stroke. These scores are used to identify uncertain pseudo-label positions
during CMW-ATC student training.

\subsection{Confidence-Guided Markov-Weighted Alternative Temporal Classification}
\label{sec:cmwatc}
CMW-ATC uses the stroke-level confidence score to determine how each
pseudo-label position is treated during training. If $c_u\geq\tau$, the
corresponding pseudo-label is treated as reliable and kept fixed, although it
is not regarded as ground truth. If $c_u<\tau$, the position is treated as
uncertain and all strokes in $\mathcal{S}$ are allowed as alternatives.
Similar to the replacement form of Alternative Temporal Classification
(ATC-R)~\cite{ATC}, CMW-ATC allows alternative labels at uncertain positions.
However, unlike ATC-R, CMW-ATC uses reliable neighbouring pseudo-labels and
learned stroke-transition probabilities to weight these alternative sequences
differently.

\subsubsection{Stroke Transition Model}
\label{sec:transition_model}

We estimate a first-order Markov model from the original labelled training
sequences. If $N(s_i,s_j)$ is the number of times $s_j$ follows $s_i$, add-one
smoothing gives
\begin{equation}
P(s_j\mid s_i)
=
\frac{N(s_i,s_j)+1}
{\displaystyle\sum_{s'\in\mathcal{S}}N(s_i,s')+K}.
\label{eq:transition_probability}
\end{equation}
Pseudo-labels and validation
or test sequences are not used. The resulting transition matrix $P$ remains fixed during student training.
We also estimate the initial-state distribution from training data using add-one smoothing: 
\begin{equation}
\rho(s_j)
=
\frac{N_{\mathrm{first}}(s_j)+ 1}
{\displaystyle\sum_{k=1}^{K}N_{\mathrm{first}}(s_k)+K},
\label{eq:initial_state_probability}
\end{equation}
where $N_{\mathrm{first}}(s_j)$ counts labelled sequences whose first stroke is
$s_j$ .

\subsubsection{Markov Weighting of Uncertain Stroke Sequences}
\label{sec:markov_uncertain_sequences}
Consecutive low-confidence positions are grouped into maximal uncertain spans.
Let $[a,b]$ denote such a span and let $m=b-a+1$ denote its length. Thus, $c_u<\tau$ for all $u\in[a,b]$, while the neighbouring positions,
when present, are reliable. For a single-position span, the corresponding empty product is taken as one and the empty sum as zero.

The first-order Markov model defined in
Section~\ref{sec:transition_model} assigns the probability of a sequence of
$m$ strokes $z_{1:m}=[z_1,\ldots,z_m]$ as
\begin{equation}
P(z_{1:m})
=
\rho(z_1)
\prod_{i=2}^{m}P(z_i\mid z_{i-1}),
\label{eq:markov_sequence_probability}
\end{equation}
where $\rho$ is the initial-state distribution and $P(z_i\mid z_{i-1})$
is the learned stroke-transition probability. However, pseudo-label sequences
generated by the teacher are not necessarily reliable at every position.
When low-confidence positions occur consecutively, their predicted strokes
may not be correct.

Candidate sequences within an uncertain span are weighted using the
stroke-transition probabilities and initial-state distribution defined in
Section~\ref{sec:transition_model}.
The weighting of an uncertain span depends on the available reliable
neighbouring pseudo-labels. A reliable boundary may be available on both sides, only on the left, only
on the right, or on neither side. These four cases determine how the
candidate sequences are weighted.
When reliable pseudo-labels are available on both sides of the span
($a>1$ and $b<\hat{U}$), let
$\ell=\hat{y}_{a-1}$ and $r=\hat{y}_{b+1}$ denote the reliable left and right
boundary strokes, respectively. 

For a candidate sequence
$z_{a:b}=[z_a,\ldots,z_b]$, the probability of the path from $\ell$ through
the candidate sequence to $r$ is
\begin{equation}
P(z_{a:b},r\mid\ell)
=
P(z_a\mid\ell)
\prod_{i=a+1}^{b}P(z_i\mid z_{i-1})
P(r\mid z_b).
\label{eq:markov_path_probability}
\end{equation}
Since this path contains $m+1$ transitions from $\ell$ to $r$, summing its
probability over all possible length-$m$ candidate sequences gives
$(P^{m+1})_{\ell r}$. Conditioning on the two reliable boundaries therefore gives the
Markov bridge distribution
\begin{equation}
q_{\mathrm{BB}}(z_{a:b}\mid\ell,r)
=
\frac{
P(z_a\mid\ell)
\displaystyle\prod_{i=a+1}^{b}P(z_i\mid z_{i-1})
P(r\mid z_b)
}{
(P^{m+1})_{\ell r}
}.
\label{eq:markov_bridge}
\end{equation}
If the uncertain span reaches the end of the pseudo-label sequence
($a>1$ and $b=\hat{U}$), the reliable left boundary
$\ell=\hat{y}_{a-1}$ remains available, but there is no right boundary.
The candidate distribution is therefore
\begin{equation}
q_{\mathrm{L}}(z_{a:b}\mid\ell)
=
P(z_a\mid\ell)
\prod_{i=a+1}^{b}P(z_i\mid z_{i-1}).
\label{eq:left_conditioned_law}
\end{equation}
Because the transition matrix is row-normalized, summing this expression over
all length-$m$ candidate sequences gives one, so no additional normalization
term is required.

Conversely, if the uncertain span begins at the first pseudo-label position
($a=1$ and $b<\hat{U}$), no reliable left boundary is available. The first candidate stroke is weighted by the initial-state distribution,
while the reliable right boundary $r=\hat{y}_{b+1}$ is included in the
path probability:
\begin{equation}
q_{\mathrm{R}}(z_{a:b}\mid r)
=
\frac{
\rho(z_a)
\displaystyle\prod_{i=a+1}^{b}P(z_i\mid z_{i-1})
P(r\mid z_b)
}{
(\rho P^{m})_{r}
}.
\label{eq:right_conditioned_law}
\end{equation}
The denominator $(\rho P^{m})_{r}$ is the total probability of reaching
$r$ after the $m$ candidate positions and normalizes the distribution over
all possible candidate sequences.

Finally, if the uncertain span covers the entire pseudo-label sequence
($a=1$ and $b=\hat{U}$), neither boundary is available. The candidate distribution then reduces to the initial-state Markov
distribution:
\begin{equation}
q_{\varnothing}(z_{1:m})
=
\rho(z_1)
\prod_{i=2}^{m}P(z_i\mid z_{i-1}).
\label{eq:unconditional_path_law}
\end{equation}
Since both $\rho$ and the transition matrix are normalized,
$q_{\varnothing}$ also sums to one over all length-$m$ candidate sequences.
Thus, $q_{\mathrm{BB}}$, $q_{\mathrm{L}}$, $q_{\mathrm{R}}$, and
$q_{\varnothing}$ covers all possible boundary cases of an uncertain span,
while using reliable neighbouring pseudo-labels when available.

\subsubsection{Candidate Graph Construction}
\label{sec:graph_construction}
For each pseudo-label position $u$, define the candidate-label set
\begin{equation}
\mathcal{C}_u
=
\begin{cases}
\{\hat{y}_u\}, & c_u\geq\tau,\\
\mathcal{S}, & c_u<\tau.
\end{cases}
\label{eq:candidate_label_set}
\end{equation}
A candidate sequence is denoted by
$Z=[z_1,\ldots,z_{\hat{U}}]$. The candidate space is therefore
\begin{equation}
\mathcal{Z}
=
\mathcal{C}_1
\times\mathcal{C}_2
\times\cdots\times
\mathcal{C}_{\hat{U}}.
\label{eq:candidate_space}
\end{equation}
The four boundary cases derived in
Section~\ref{sec:markov_uncertain_sequences} can be represented in a single graph form.
For an uncertain span $[a,b]$, the left and right boundary factors
$g_L(z_a)$ and $g_R(z_b)$ are defined based on whether reliable boundary
strokes are available:
\begin{equation}
\begin{aligned}
g_L(z_a)
&=
\begin{cases}
P(z_a\mid\ell), & \ell\text{ available},\\
\rho(z_a), & \ell\text{ unavailable},
\end{cases}
\\[4pt]
g_R(z_b)
&=
\begin{cases}
P(r\mid z_b), & r\text{ available},\\
1, & r\text{ unavailable}.
\end{cases}
\end{aligned}
\label{eq:boundary_factors}
\end{equation}
where $\ell=\hat{y}_{a-1}$ and $r=\hat{y}_{b+1}$ denote the reliable strokes
before and after the span, respectively, when available. The Markov log-score of a candidate sequence is then defined as
\begin{equation}
\begin{aligned}
A_{\mathrm{CMW}}(z_{a:b})
&=
\log g_L(z_a)
+
\sum_{i=a+1}^{b}
\log P(z_i\mid z_{i-1})\\
&\quad
+
\log g_R(z_b).
\end{aligned}
\label{eq:cmg_log_score}
\end{equation}
This score combines the left boundary factor, the transitions between candidate strokes within the span, and the right boundary factor. The candidate graph preserves the length of the teacher pseudo-label sequence
and allows alternative stroke labels only at uncertain positions.

\subsubsection{CMW-ATC Training Objective}
\label{sec:cmgatc_objective}
For an uncertain span $[a,b]$, the Markov scores from
Eq.~\eqref{eq:cmg_log_score} are normalized over all candidate sequences to
obtain their weights:
\begin{equation}
q(z_{a:b})
=
\frac{\exp A_{\mathrm{CMW}}(z_{a:b})}
{\displaystyle
\sum_{\tilde{z}_{a:b}\in\mathcal{Z}_{a:b}}
\exp A_{\mathrm{CMW}}(\tilde{z}_{a:b})},
\label{eq:candidate_weight}
\end{equation}
where $\mathcal{Z}_{a:b}=\mathcal{C}_a\times\cdots\times\mathcal{C}_b$ denotes
the candidate space for the span. Depending on the boundary case, this weight corresponds to
$q_{\mathrm{BB}}$, $q_{\mathrm{L}}$, $q_{\mathrm{R}}$, or
$q_{\varnothing}$ defined in
Section~\ref{sec:markov_uncertain_sequences}.
If a sequence contains $J$ separated uncertain spans, the weight of a
candidate sequence is
\begin{equation}
q_{\mathrm{CMW}}(Z)
=
\prod_{j=1}^{J}
q(z_{a_j:b_j}),
\label{eq:sequence_graph_weight}
\end{equation}
where $[a_j,b_j]$ denotes the $j$-th uncertain span. When $J=0$, the only
candidate sequence is the teacher prediction, and its weight is one. Since each uncertain-span distribution is normalized, the sequence weights
satisfy
$\sum_{Z\in\mathcal{Z}} q_{\mathrm{CMW}}(Z)=1$.

The student model assigns a CTC probability to each candidate sequence.
The loss for unlabelled data is therefore
\begin{equation}
\begin{aligned}
\mathcal{L}_{\mathrm{CMW\text{-}ATC}}(X,\hat{Y})
&=
-\log
\sum_{Z\in\mathcal{Z}}
q_{\mathrm{CMW}}(Z)
P_{\theta_S}^{\mathrm{CTC}}(Z\mid X).
\end{aligned}
\label{eq:cmg_atc_loss}
\end{equation}
The weighted sum over candidate sequences is computed on the candidate graph
without explicitly enumerating all candidate sequences.
The overall student loss is
\begin{equation}
\mathcal{L}_{\mathrm{total}}
=
\mathcal{L}_{\mathrm{CTC}}(X_L,Y_L)
+
\lambda
\mathcal{L}_{\mathrm{CMW\text{-}ATC}}(X_U,\hat{Y}_U),
\label{eq:total_loss}
\end{equation}
where $\lambda$ controls the contribution of the unlabelled data.

\section{Experimental Setup}
\label{sec:exp_setup}
This section describes the datasets, model configuration, training procedure,
baselines, ablation studies, and hyperparameter settings used for evaluation.

\subsection{Datasets}
We use four source datasets for Tabla Stroke Transcription (TST): Tabla
Identification Dataset (TID), Tabla Solo Dataset (TSD), Hindustani Music
Rhythm Dataset (HMR), and TID-S. TID and TID-S were introduced in~\cite{ADRM}, while TSD~\cite{S8} and HMR~\cite{G4} are
publicly available. These are organized into three evaluation sets:
\textit{TID+TSD}, \textit{HMR}, and \textit{TID-S}. TSD is combined with
TID because it contains limited labelled training data. The three settings
cover tabla solo recordings, Hindustani music with vocal and instrumental
accompaniment, and synthetic tabla recordings with controlled rhythmic
variation. Their training, validation, and test partitions are summarized
in Table~\ref{T_dataset}, with further dataset details provided in
Supplementary Material (S.I).

The labelled training partitions are augmented following~\cite{S9} using
pitch shifting, time scaling, attack remixing, spectral filtering, and
stroke remixing; validation and test data are not augmented. For semi-supervised learning, we additionally
use an unlabelled YouTube corpus containing 676 minutes of tabla solo
performances and 602 minutes of Hindustani concert recordings. The same
unlabelled corpus is used for all semi-supervised methods.

\subsection{Feature Extraction and Model Architecture}
All experiments use 128-dimensional log-Mel spectrograms with a
2048-sample analysis window and a 10~ms frame shift at 44.1~kHz.
The features are mean-normalized per frequency. The teacher and student use the C-TDNN-F architecture adopted for
TST in~\cite{ADRM}, based on factorized TDNNs~\cite{Povey2018} and
convolutional TDNN-F models~\cite{Psutka2021}. The model consists of 12 C-TDNN-F layers
with bottleneck projections, temporal convolution with context
$[-1,0,+1]$, residual connections, batch normalization, and dropout (0.1),
followed by a linear projection and CTC output layer. Increasing dilation
factors are used to expand the temporal receptive field. The same
architecture is used for all learning-based semi-supervised methods.

\subsubsection{Confidence Estimation Model}

S-CEM adapts the CEM architecture used in the CTC-based TruCLeS
framework~\cite{ravi2025asr} to stroke-level confidence estimation, with
three fully connected layers of 512, 256, and 128 units, ReLU hidden
activations, and a sigmoid output. It is trained using Adam with a learning
rate of $10^{-3}$ for 100 epochs. S-CEM is kept fixed during student training.

\subsection{Semi-Supervised Training Procedure}
A separate CTC teacher is trained for each evaluation setting using its
corresponding labelled training data. After training, each teacher is frozen and used
to obtain the predictions and representations required for S-CEM training and
to generate pseudo-labels for the unlabelled recordings.
For each evaluation setting, $P$ and $\rho$ are estimated independently
from its original labelled training sequences and remain fixed during
student training. Augmented sequences, pseudo-labels, validation data, and
test data are not used for their estimation.
A separate student is trained from scratch for each evaluation setting using
conventional CTC on labelled recordings and CMW-ATC on pseudo-labelled
recordings.

\subsection{Baseline Methods}

All methods are evaluated as symbolic stroke sequences using the same
sequence-level Stroke Error Rate (SER). We consider the onset-level
supervised OTM~\cite{S10} and Segment-and-Classify (S\&C)~\cite{S12}
methods, together with the stroke-clustering framework
in~\cite{1_4_gowriprasad2025unsupervised}. OTM and S\&C are evaluated on
HMR, while clustering is evaluated on all three evaluation settings. S\&C
retains its onset-based segmentation and is adapted to the 30-stroke
vocabulary. The clustering method does not use stroke labels
during cluster formation; labelled training data are used only to map the
resulting clusters to the target stroke vocabulary.

For semi-supervised learning, we consider a conventional CTC
teacher--student baseline and ATC-R~\cite{ATC}. The teacher--student baseline
trains on teacher-generated pseudo-labels using conventional CTC. ATC-R retains its Contrastive-CTC training, CTC-based token confidence,
automatic thresholding, and scaled wildcard replacement. Unlike the original ATC-R framework, the teacher is kept
fixed during student training to match the frozen-teacher protocol used by
CMW-ATC. Additional baseline details
are provided in Supplementary Sections S.II and S.III.

\subsection{Ablation Studies}
The ablation study examines the two main components of the proposed framework:
stroke-level confidence estimation and Markov-based candidate weighting.
ATC-R~\cite{ATC} is included as the reference alternative-label method.
A1 combines S-CEM confidence with the scaled wildcard replacement of ATC-R.
A2 uses S-CEM with all-stroke candidates and assigns equal weight to all
candidate sequences, without Markov weighting. A3 retains the same confidence and candidates as A2, but applies
forward Markov weighting using the reliable left boundary, when available,
and the within-span transitions, without using the right boundary.

The proposed CMW-ATC additionally uses the reliable right boundary when
available. Thus,
A2 versus A3 evaluates the contribution of Markov transition weighting, while
A3 versus CMW-ATC evaluates the contribution of the right reliable boundary.
To separately evaluate confidence estimation, A4 uses CTC-based token
confidence with the same  CMW weighting. Therefore, A4 versus the
proposed system evaluates the contribution of S-CEM while keeping the
candidate formulation and Markov weighting unchanged.

\subsection{Hyperparameter Settings}
CMW-ATC has two task-specific hyperparameters: the confidence threshold
$\tau$ and the pseudo-label loss weight $\lambda$. The transition matrix
$P$ and initial-state distribution $\rho$ use fixed add-one smoothing and
have no tunable hyperparameters. The teacher and student models are trained
using Adam with a learning rate of $10^{-3}$ for 100 epochs. The confidence
threshold and loss weight are selected jointly on the corresponding
validation set from $\tau\in\{0.4,0.5,\ldots,0.9\}$ and
$\lambda\in\{0.1,0.2,\ldots,1.0\}$. The selected values are
$(\tau,\lambda)=(0.6,0.5)$ for TID+TSD and TID-S, and
$(\tau,\lambda)=(0.5,0.4)$ for HMR.

For ATC-R, the wildcard scale is tuned on the validation set to $\eta=0.4$ and then held fixed for test evaluation; its original automatic confidence-thresholding procedure is retained. For A4, the
CTC-based confidence threshold is selected on the corresponding validation
set from the same threshold range used for S-CEM. The remaining ablations
use the corresponding CMW-ATC hyperparameters. Decoding beam-search parameters are
$k_{\mathrm{beam}}=120$ and $\Delta_{\mathrm{beam}}=10$, while CMW
candidate weighting is computed on the full candidate graph without beam
pruning.

\begin{table*}[t]
\centering
\begin{minipage}[t]{0.24\textwidth}
\centering
\caption{Dataset Statistics}
\label{T_dataset}
\setlength{\tabcolsep}{2pt}
\begin{tabular}{@{}lccc@{}}
\toprule
\textbf{Dataset} &
\makecell[c]{\textbf{Train}\\\textbf{(min)}} &
\makecell[c]{\textbf{Val.}\\\textbf{(min)}} &
\makecell[c]{\textbf{Test}\\\textbf{(min)}} \\
\midrule
\textbf{TID+TSD} & 103+412* & 24 & 24 \\
\textbf{HMR}     & 212       & 60 & 30 \\
\textbf{TID-S}   & 81+324*   & 20 & 20 \\
\bottomrule
\end{tabular}
\begin{tablenotes}
\item * indicates augmented data.
\end{tablenotes}
\end{minipage}
\hfill
\begin{minipage}[t]{0.73\textwidth}
\centering
\caption{Stroke Error Rate (SER, \%) across TST systems.}
\vspace{-2mm}
\label{tab:main_results}
\setlength{\tabcolsep}{1.5pt}

\begin{tabular}{@{}l cc c c cccc@{}}
\toprule
\textbf{Data} &
\multicolumn{2}{c}{\textbf{Strongly Supervised}} &
\textbf{Clustering} &
\multicolumn{4}{c}{\textbf{Sequence-level / Semi-Supervised}} \\
\cmidrule(lr){2-3}
\cmidrule(lr){4-4}
\cmidrule(lr){5-8}
&
\textbf{OTM~\cite{S10}} &
\textbf{S\&C~\cite{S12}} &
\textbf{Based~\cite{1_4_gowriprasad2025unsupervised}} &
\textbf{CTC} &
\textbf{Teacher--Student} &
\textbf{ATC-R \cite{ATC}} &
\textbf{Proposed CMW-ATC}\\
\midrule

\textbf{TID+TSD} &
-- &
-- &
48.2&
25.1 &
24.2 ($\downarrow$3.6\%) &
22.9 ($\downarrow$8.8\%) &
\textbf{18.3 ($\downarrow$27.1\%)} \\

\textbf{HMR} &
28.5 &
67.4 &
84.1 &
43.8 &
42.7 ($\downarrow$2.5\%) &
39.6 ($\downarrow$9.6\%) &
\textbf{33.3 ($\downarrow$24.0\%)} \\

\textbf{TID-S} &
-- &
-- &
42.3 &
17.8 &
16.7 ($\downarrow$6.2\%) &
16.0 ($\downarrow$10.1\%) &
\textbf{12.7 ($\downarrow$28.7\%)} \\

\bottomrule
\end{tabular}
\begin{tablenotes}
\item Parenthesized values denote relative SER reduction ($\downarrow$)
relative to the CTC (Labelled Data) baseline.
\item ``--'' denotes not applicable, since the required onset-level
annotations for training the corresponding baselines are unavailable for
the dataset.
\end{tablenotes}
\end{minipage}
\end{table*}

\begin{table*}[t]
\centering
\caption{Ablation study of confidence estimation and candidate weighting
in the proposed CMW-ATC framework.}
\label{tab:ablation}

\small
\renewcommand{\arraystretch}{0.95}

\begin{tabular*}{\textwidth}{
@{\extracolsep{\fill}}
l
c
c
ccc
@{}
}
\toprule
\textbf{Method} &
\textbf{Confidence} &
\textbf{Candidate formulation / weighting} &
\textbf{TID+TSD} &
\textbf{HMR} &
\textbf{TID-S} \\
\midrule

ATC-R \cite{ATC} &
CTC-based &
Wildcard $*$ &
22.9 & 39.6 & 16.0 \\

A1 &
S-CEM &
Wildcard $*$ &
21.4 & 37.8 & 14.8 \\

A2 &
S-CEM &
All strokes -- Uniform weighting &
21.2 & 38.1 & 14.6 \\

A3 &
S-CEM &
All strokes -- Forward Markov weighting &
20.1 & 35.9 & 13.8 \\

A4 &
CTC-based &
All strokes -- CMW weighting &
20.7 & 36.9 & 14.2 \\

Proposed CMW-ATC &
S-CEM &
All strokes -- CMW weighting &
\textbf{18.3} & \textbf{33.3} & \textbf{12.7} \\

\bottomrule
\end{tabular*}
\end{table*}

\section{Results and Discussion}
\label{sec:results}
Performance is evaluated using Stroke Error Rate (SER), defined as
\begin{equation}
\mathrm{SER}(\%)=
\frac{S+D+I}{N}\times100,
\label{eq:ser}
\end{equation}
where $S$, $D$, and $I$ denote the numbers of substitution, deletion, and
insertion errors, respectively, and $N$ is the number of reference strokes.
Lower SER indicates better transcription performance. Unless
otherwise stated, all results are reported on the corresponding test sets.

Table~\ref{tab:main_results} presents the transcription results across the
evaluated TST systems. The proposed CMW-ATC framework obtains the lowest SER
among the sequence-level and semi-supervised methods on all three evaluation
sets. Relative to the labelled-data CTC baseline, CMW-ATC reduces SER by
24.0--28.7\%, while reducing SER relative to ATC-R by 15.9--20.6\%.
Teacher--Student improves over the labelled-data CTC baseline, while ATC-R
provides a further reduction in SER. 

Compared with the other TST approaches, CMW-ATC obtains lower SER than the
clustering baseline on all three evaluation sets and lower SER than the
onset-based S\&C method on HMR, while requiring only sequence-level stroke
annotations. A key methodological distinction is that CMW-ATC performs
sequence-level transcription with CTC and uses learned stroke-transition
information to weight alternative stroke candidates at uncertain positions.
In contrast, the clustering method first groups acoustic segments and maps
the resulting clusters to stroke labels, while S\&C classifies detected
strokes from local acoustic segments. OTM achieves a lower SER than CMW-ATC
on HMR, but requires onset-level annotations for training. Thus, CMW-ATC
achieves the lowest SER among the evaluated sequence-level and
clustering-based methods without requiring frame- or onset-level stroke
annotations.

HMR is the most challenging evaluation setting, with recordings from 45
artists and vocal and instrumental accompaniment across four
$t\bar{a}las$. The higher SER observed on HMR may be associated with this
greater acoustic and performer variability; nevertheless, CMW-ATC maintains
the lowest SER among the evaluated sequence-level and semi-supervised
methods.

Table~\ref{tab:ablation} summarizes the ablation studies on confidence
estimation and candidate weighting. Under the wildcard formulation, replacing
CTC-based token confidence with S-CEM (ATC-R versus A1) reduces SER by
4.5--7.5\% relative across the three evaluation sets. A4 and CMW-ATC provide a controlled comparison because they use the same set of all-stroke candidates and the same CMW weighting. Replacing CTC-based
confidence with S-CEM in this setting reduces SER by 9.8--11.6\%, showing
the contribution of the proposed confidence estimator.

The effect of candidate weighting is evaluated through A2, A3, and the
proposed CMW-ATC, which use the same S-CEM confidence and all-stroke
candidates. Replacing uniform weighting in A2 with forward Markov weighting
in A3 reduces SER by 5.2--5.8\% relative. Adding the reliable right-boundary transition in CMW-ATC further reduces SER by 7.2--9.0\% relative to A3. These results show that both Markov transition weighting and right-boundary information improve performance.

\begin{figure*}[t]
\centering
\includegraphics[width=\textwidth]{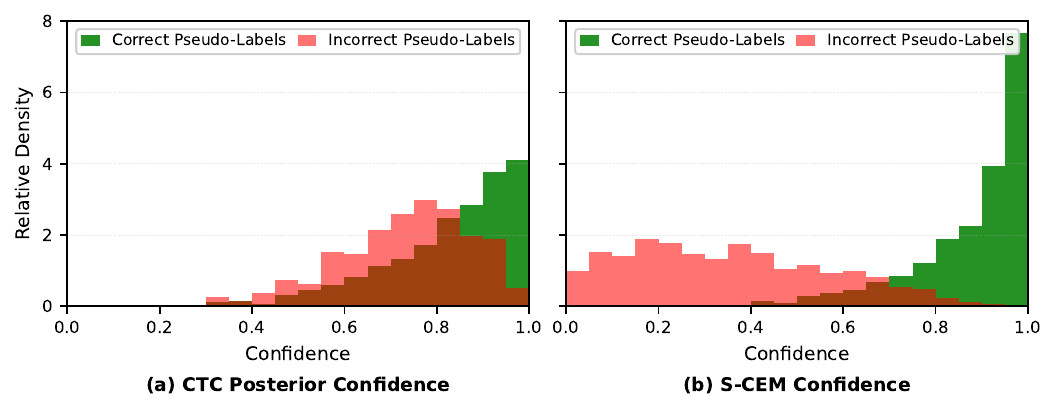}
\vspace{-0.5cm}
\caption{Confidence distributions of correct and incorrect teacher-predicted
strokes on the TID+TSD test set.}
\label{fig:confidence}
\end{figure*}

Fig.~\ref{fig:confidence} shows the confidence distributions of correct and
incorrect teacher-predicted strokes on TID+TSD. With CTC posterior confidence, the distributions of correct and incorrect
predictions overlap considerably. In contrast, S-CEM gives higher confidence mainly to
correct predictions and lower confidence to incorrect predictions, producing
a clearer separation between the two distributions.

\subsection{Transcription Error Analysis}
\label{sec:error_analysis}

To further analyze the transcription results, we decompose the errors into
substitutions (S), deletions (D), and insertions (I). Table~\ref{tab:overall_sdi}
shows that substitutions remain the largest error category for CMW-ATC across
all three evaluation sets, while all three error types are reduced compared
with the CTC baseline. This indicates that the proposed approach improves
transcription performance across different types of sequence errors, with
substitution errors remaining the primary source of residual errors.

The largest relative reduction is observed for insertion errors, with
reductions of 38.9\%, 41.5\%, and 44.4\% on TID+TSD, HMR, and TID-S,
respectively. The candidate construction, however, preserves the number of
pseudo-label positions and does not explicitly add or remove stroke
positions. The reductions in insertion and deletion errors should therefore
be viewed as effects of the resulting student model rather than as direct
corrections performed by the candidate graph. CMW-ATC instead focuses on
handling uncertainty in stroke identity at low-confidence positions, which
contributes directly to the reduction in substitution errors. Nevertheless,
substitution errors remain the dominant error category across all three
evaluation sets.

\begin{table}[!t]
\centering
\caption{Overall error distribution for TID+TSD, HMR, and TID-S. Values in parentheses denote the relative contribution of each error type to SER.}
\label{tab:overall_sdi}

\renewcommand{\arraystretch}{1.0}
\setlength{\tabcolsep}{2.0pt}

\begin{tabular}{llcccc}

\noalign{\vskip 1pt}
\hline
\noalign{\vskip 1.5pt}

Dataset & Method & SER & S & D & I \\

\noalign{\vskip 1.5pt}
\hline
\noalign{\vskip 1pt}

TID+TSD & CTC
& 25.1 & 15.2 (60.6\%) & 4.5 (17.9\%) & 5.4 (21.5\%) \\

& CMW-ATC
& 18.3 & 11.5 (62.9\%) & 3.4 (18.8\%) & 3.3 (18.3\%) \\

\noalign{\vskip 2pt}
\hline
\noalign{\vskip 2pt}

HMR & CTC
& 43.8 & 22.4 (51.1\%) & 12.0 (27.4\%) & 9.4 (21.5\%) \\

& CMW-ATC
& 33.3 & 18.9 (56.7\%) & 9.0 (26.9\%) & 5.5 (16.4\%) \\

\noalign{\vskip 2pt}
\hline
\noalign{\vskip 2pt}

TID-S & CTC
& 17.8 & 10.8 (60.7\%) & 3.4 (19.1\%) & 3.6 (20.2\%) \\

& CMW-ATC
& 12.7 & 8.3 (65.7\%) & 2.4 (18.7\%) & 2.0 (15.7\%) \\

\noalign{\vskip 1pt}
\hline

\end{tabular}
\end{table}

\section{Conclusion}
\label{sec:conclusion}

We presented a semi-supervised framework for Tabla Stroke Transcription that
combines stroke-level confidence estimation with Confidence-Guided Markov
Weighting Alternative Temporal Classification (CMW-ATC). S-CEM identifies
uncertain teacher-predicted strokes, while CMW-ATC represents the corresponding
uncertain spans using alternative candidate sequences and weights them using
learned stroke-transition probabilities and reliable boundary information.

The proposed system obtains SERs of 18.3, 33.3, and 12.7 on TID+TSD, HMR,
and TID-S, respectively, corresponding to relative SER reductions of
24.0--28.7\% over the labelled-data CTC baseline and 15.9--20.6\% over
ATC-R. The ablation studies show contributions from S-CEM, forward Markov
weighting, and the additional use of reliable right-boundary information in
the complete CMW formulation. Error analysis shows reductions in
substitution, deletion, and insertion errors, although CMW-ATC directly
handles uncertainty in stroke identity and does not explicitly model
insertion or deletion errors. Future work will investigate broader
low-resource TST settings and further reductions in annotation requirements.

\bibliographystyle{IEEEtran}
\bibliography{main}

\clearpage

\vfill

\end{document}